\documentclass{ws-ijmpa}
\usepackage[super,compress]{cite}
\usepackage{braket}

\begin{document}

\markboth{Richard Herrmann}{Covariance in Fractional Calculus}

\catchline{}{}{}{}{}

\title{Covariance in Fractional Calculus}

\author{Richard Herrmann}

\address{Berliner Ring 80, 63303 Dreieich, Germany\\
r.herrmann@fractionalcalculus.org}

\maketitle

\begin{history}
\received{Day Month Year}
\revised{Day Month Year}
\end{history}

\begin{abstract}
Based on the requirement of covariance, we propose a new approach for generalizing fractional calculus in multi-dimensional space. 
As a first application we calculate an approximation for the ground state energy of the fractional two-dimensional harmonic oscillator using the Ritz variational principle.
 In lowest order perturbation theory a  new analytic result is derived for the energy level spectrum of the one-dimensional fractional harmonic oscillator.
\end{abstract}

\ccode{PACS numbers:}
04.30.-w; 05.40.+j;   31.15.xt
\section{Introduction}
The first applications of fractional calculus were developed in 1-dimensional Cartesian space, where problems were defined using a single coordinate:

As an example, Abel's treatment of the tautochrone problem used the path-length $s$ \cite{abel}.

In a similar way, as long as fractional calculus is understood as a framework for investigating  memory effects such as investigating the influence of history on current developments \cite{pod99, hil00}, implementing causal and anti-causal properties or describing phenomena like anomalous diffusion processes \cite{met00} or causal elastic waves \cite{nas13} these processes are time-dependent and can be treated using a single time coordinate $t$.

While there is a long tradition in the theory of fractional calculus on $R^n$ \cite{sam93, pod99}, practical interest in the multi-dimensional generalization of fractional calculus in Cartesian space, particularly for fractional wave equations, began with Raspini's work on the fractional Dirac equation, which introduced derivatives of order $\alpha=2/3$ \cite{ras00}, and has recently gained increasing interest  \cite{tar21, tar23, kos24}.

When extending fractional calculus to higher-dimensional spaces, ensuring covariance is essential. This requires that the definition of the fractional derivative and the corresponding fractional differential equations remain invariant under arbitrary coordinate transformations which is fundamental to the general validity of the results derived \cite{mis73, lee18}.

\section{The starting point in one dimensional Cartesian coordinates}
Fractional calculus 
extends the concept of a derivative operator from integer order $n$ to arbitrary order $\alpha$, where $\alpha$ is a real or complex value
\begin{equation}
\label{c1first}
{d^n \over dx^n} \rightarrow {d^\alpha \over dx^\alpha}, \qquad n \in \mathbb{N}, \alpha \in \mathbb{C}.
\end{equation}
This generalization introduces the concept of non-locality to what was previously a local derivative operator, realized through convolution integrals with weakly singular kernels or weights.

The definition of a fractional order derivative is not unique,  several definitions 
e.g. the Riemann \cite{rie47}, Caputo\cite{cap67}, Riesz  \cite{rie49} coexist and seem equally well suited for an extension of the standard derivative.

For example, in the one dimensional case the left sided, causal Riemann fractional derivative is given by:
\begin{equation}
D_R^\alpha f(x) =  \frac{1}{\Gamma(1-\alpha)} \int_0^x dh  \frac{1}{h^\alpha} f'(x-h), \qquad x \geq 0, \quad 0 \leq \alpha \leq 1   
\end{equation}
with $f'(x) = \frac{d}{dx}f(x)$ and $\Gamma$ the Euler gamma function.   

The  left sided, causal Caputo fractional derivative is given by:
\begin{equation}
D_C^\alpha f(x) = \frac{1}{\Gamma(1-\alpha)} \frac{d}{dx} \int_0^x dh  \frac{1}{h^\alpha} f(x-h) , 
\qquad x \geq 0, \quad 0 \leq \alpha \leq 1   
\end{equation}
Both definitions differ in the sequence of two operations: The Riemann fractional derivative first applies a standard derivative on $f(x)$ followed by a fractional integral. The Caputo fractional derivative follows the reversed order, first applying a fractional integral and then a standard derivative.

The Riesz fractional derivative extends the second derivative $\frac{d^2}{dx^2}$ to the fractional case and is defined as
\begin{eqnarray}
\label{q12driesz}
^\infty D^\alpha_\textrm{\tiny{RZ}} f(x)\! \! &=&\! \!  \Gamma(1+\alpha)
{\sin(\pi \alpha/2)\over \pi} 
 \int_0^\infty dh \, {f(x+h)-2 f(x) + f(x-h) \over h^{\alpha+1}} \\
& &  \qquad  \qquad \qquad  \qquad \qquad \qquad  \qquad \qquad  \qquad  \qquad  0< \alpha <2 \nonumber
\end{eqnarray}
where the left superscript in $^\infty D^\alpha_\textrm{\tiny{RZ}}$ emphasizes that the Riesz fractional derivative is a superposition of left and right sided fractional derivatives,  resulting in an integral domain that actually spans the full space $\mathbb{R}$.

Although distinct in their action on different function classes, all definitions follow a two-step procedure: 

The classical local operator $L = \frac{d^n}{dx^n}$ is extended to the fractional, 
non-local operator $\tilde{\mathcal{L}}^\alpha_{\textrm{type}} = D^\alpha_{\textrm{type}}$ by applying a convolution 
integral with a weakly singular weight $w(h)=h^{-\alpha}$. 

In the following, we will present a  covariant procedure applicable to any local tensor operator on multi-dimensional Riemannian spaces.
\section{Covariant transition from local to non-local operators}
We define the requirement of covariance as form invariance under coordinate transformations, which on $\mathbb{R}^N$  leave the line element $ds^2$ invariant \cite{adl75}:
\begin{equation}
ds^2 = g_{i j}dx^i dx^j  \qquad i,j = 1, ..., N  
\end{equation}
and the components of the metric tensor $g_{i j}$ transform as a tensor of rank 2:
\begin{equation}
  \label{chxxgmn}
g_{k l} = \frac{\partial x^i}{\partial x^k}
\frac{\partial x^j}{\partial x^l} g_{i j}
\qquad \qquad  i,j,k,l = 1, ..., N  
\end{equation}
In the following, we will present a covariant 2-step procedure to extend an arbitrary local operator to the fractional case by applying a convolution with a covariant weight. 

We compose a general covariant fractional operator $\tilde{\mathcal{O}}_{\textrm{type}}(w)$ 
as a combination of two covariant components, the classic local operator $L$ and 
a non-localization operator $\tilde{\mathcal{G}}(w)$ with covariant weight $w$:
\begin{itemize}
\item
The classic local operator $L$ is 
the basic constituent of almost all classic field theories of physics e.g.
classic particle physics (Hamiltonian H), electro-dynamics (Maxwell equations), 
quantum mechanics (Schr\"odinger-, Klein-Gordon-, Dirac- equation), gauge theories (Yang-Mills theory) 
or cosmology (Einstein field equations, string-theories) assumed to transform as a tensor of given rank. 
\item
The non-localization operator $\tilde{\mathcal{G}}(w)$ is 
given as a convolution integral with a covariant weight $w$.
\begin{equation}
  \tilde{\mathcal{G}}(w) \ast f(\vec{x}, \vec{h}) = 
  \frac{1}{N}\int_{\mathbb{R}^N} dV(h) \, w(h) \cdot f(\vec{x}, \vec{h})
\end{equation}
with $\ast$ being a short hand notation for the convolution, with the invariant volume element $dV = d^n h \, \sqrt{g}$ and a norm $N$. 
\end{itemize}
Combining these two operators in a 2 step transition procedure from local to fractional operators   
may be realized as two different operator sequences: 

A Riemann-type sequence,  where 
first the non-localization operator is applied followed by the classic operator:
  \begin{equation}
    \label{ch245R}
    \tilde{\mathcal{L}}_{\textrm{R}}(w)  =  
    L  \cdot
    \tilde{\mathcal{G}}(w) \ast
    \end{equation}

A Caputo-type sequence with inverted operator succession, 
where first the classic operator is applied followed by the non-localization operator:
  \begin{equation}
    \label{ch245C}
    \tilde{\mathcal{L}}_{\textrm{C}}(w)   =  
     \tilde{\mathcal{G}}(w) \ast
    L  \cdot
    \end{equation}
Both types of a fractional operator will be used in the following.    

    \noindent
The local operator $L$ should transform as a tensor of a given rank $m$.

A classic example is the Laplace-operator contracted to a tensor of rank 0 (scalar) used in wave equations:
\begin{eqnarray}
\label{ch260LL}
\bigtriangleup    \Psi     & = & \nabla^{j} \nabla_{j} \Psi \\
&=& g^{ij} \nabla_{i} \nabla_{j} \Psi 
      \qquad\qquad i,j=1,...,N                                      \\
   & = & g^{ij}(\partial_{i} \partial_{j} -
\left\{\begin{array}{c}
k      \\ ij           \\
\end{array} \right\}
                 \partial_{k})    \Psi       \\ 
  & = &\frac{1}{\sqrt{g}} \partial_{i}\,
        g^{ij} \sqrt{g} \, \partial_{j} \Psi 
                                   \end{eqnarray}
where
$\left\{\begin{array}{c}
k      \\ ij           \\
\end{array} \right\}$ being the Christoffel symbol
\begin{equation}
\left\{\begin{array}{c}
k      \\ ij           \\
\end{array} \right\}
=\frac{1}{2} g^{k l}(\partial_i g_{j l}+\partial_j g_{i l}-\partial_l g_{i j})
\end{equation}
\index{Christoffel symbol}
$g$ is the determinant of the metric tensor, $g= \det g_{ij}$ (\ref{chxxgmn})
and
$\nabla$ is the covariant derivative \cite{adl75, mis73}.

\noindent
We postulate, that the fractional extension of a given local operator should not change the rank of the local operator. The fractional extension of a wave-equation should remain  a wave-equation with corresponding tensor characteristics, the fractional extension of the Fokker-Planck-equation should exhibit the same tensor properties as the local one.

Therefore  the non-localization operator  $\tilde{\mathcal{G}}(w)$ may only transform as a tensor of rank 0 (scalar). This holds  as long as the weight  $w(h)$ depends only on the line element and itself transforms 
exclusively  as a tensor of rank 0 (a scalar).

\begin{equation}
w(h) \equiv w(\sqrt{ds^2})
\end{equation}
An obvious choice, which we will use in the following, is the weakly singular power law weight used in e.g. Riesz potentials, which conserves the requirements of isotropy and homogeneity  on $\mathbb{R}^N$ \cite{tar18, die20}:  
\begin{equation}
  \label{ch24501}
  w(h) = |h|^{-\alpha} \qquad \alpha > 0 
\end{equation}

The covariant non-localization operator $\tilde{\mathcal{G}}(w)$ is realized as a convolution integral on a second independently chosen  coordinate space $h = \{h_1,h_2,...,h_n\}$ 
\begin{equation}
\label{chxxnonR}
\tilde{\mathcal{G}}(w) \ast =\frac{1}{N}  \int_{\mathbb{R}^n}  d^n h \, \sqrt{g} \, |h|^{-\alpha} \, \cdot  
\qquad \alpha > 0 
\end{equation}
We are free to choose two different coordinate systems for the local operator  $L$ and the convolution $\tilde{\mathcal{G}}$.

E.g. on $R^2$ the problem of calculating the spectrum of a rectangular membrane may be best formulated using in Cartesian coordinates $x=\{x,y\}$, but the corresponding convolution with  
Riesz weight (\ref{ch24501}) may be best solved using polar coordinates $h = \{h_r, h_\phi\}$.

In the 3-dimensional case the convolution integral (\ref{chxxnonR}) may be solved using  spherical coordinates $h = \{h_r, h_\phi, h_\theta\}$, independent of the local coordinate system used.   

We now have proposed the general strategy  to obtain 
multi-dimensional covariant fractional extensions
of classic tensor operators.

In the following we will give an example how to 
apply the presented method to obtain the fractional analogue of the classic Laplace operator.   

We will derive the fractional kinetic energy operator of 
the classic 2-dimensional Schr\"odinger equation and will then give an upper limit
for the ground state energy of the fractional Schr\"odinger equation 
with fractional rotationally symmetric power law potential using the Ritz variation principle \cite{rit09, gro88, lei05} .

\section{The Ritz variational principle and the ground state energy of the 2-dimensional fractional symmetric power law potential}
The Ritz variational principle is widely used in quantum mechanics to optimize 
parametrized wave functions by minimizing the corresponding energy expectation value \cite{rit09, gro88, lei05} 

We will demonstrate the validity of the above proposed procedure by approximating the ground state energy $E_0$ 
for the 2-dimensional fractional Schr\"odinger equation.

We will first derive the fractional extension of the classic Schr\"odinger equation
and then calculate the expectation value for a 2-dimensional Gaussian trial function.

The classic 2-dimensional Schr\"odinger equation in coordinate representation 
in natural units ($m=c=\hbar=1$) is given in Cartesian coordinates $\{x,y\}$ as the sum
of kinetic and potential energy operators: 
\begin{equation}
  \label{ch245Hold}
  H = T+V=\left( -{ 1\over 2} \Delta + V(x,y) \right) \Phi(x,y) = E \Phi(x,y) 
 \end{equation}
with the rotationally symmetric power law potential ($r^2 = x^2+y^2$) 
\begin{equation}
 V = { 1\over 2} (x^2+y^2)^{\sigma/2}= { 1\over 2} r^{\,\sigma}  
\end{equation}
which in the case $\sigma=2$ reduces to the rotationally symmetric harmonic oscillator.

Since we will use the Riesz weight, we will make the following settings for the 
operators $\{ \hat{p}_i, \hat{x}_i \}$ from the classic to the fractional case
to ensure the equivalence of coordinate and momentum representation: 
\begin{eqnarray}
  \label{ch245trans}
  \Delta = \hat{p}^2= \hat{p}_i \hat{p}^i &\rightarrow&  D_{\alpha/2} \Delta^{\alpha/2} = D_{\alpha/2}(\hat{p}_i \hat{p}^i)^{\alpha/2} \\
  \hat{x}^2 = \hat{x}_i \hat{x}^i &\rightarrow&  D_{\alpha/2} (\hat{x}^2)^{\alpha/2} = D_{\alpha/2}(\hat{x}_i \hat{x}^i)^{\alpha/2} 
  \end{eqnarray}
where $ D_{\alpha/2}$ is a dimensional factor to ensure correct kinetic and potential energy units in the fractional case. 

The fractional Schr\"odinger equation is then the extended version of classic (\ref{ch245Hold}) and 
in natural units ($m=c=\hbar= D_{\alpha/2} = D_{\beta/2} = 1$) given as the sum of the fractional extension of kinetic and potential energy in coordinate representation:
\begin{eqnarray}
  \label{ch245H}
  H^{\alpha,\beta} &=& T^{\alpha/2}+V^{\beta/2} =
  \left( -{ 1\over 2} \Delta^{\alpha/2} + V^{\beta/2}(x,y) \right) \Phi(x,y) 
  = E \Phi(x,y) \nonumber \\
  & & \qquad \qquad \qquad\qquad  \qquad \qquad\qquad \qquad\qquad 0 \leq \alpha \leq 2
 \end{eqnarray}
with the fractional rotationally symmetric power law potential ($r^2 = x^2+y^2$) 
\begin{equation}
  V^{\beta/2}(x,y) = { 1\over 2} (x^2+y^2)^{\beta/2}= { 1\over 2} r^{\,\beta} 
  \qquad \beta, r \geq 0
\end{equation}
Obviously (\ref{ch245H}) reduces to the 
fractional harmonic oscillator for the case $\alpha=\beta$ and to the 
classical Schr\"odinger equation (\ref{ch245Hold}) with the standard harmonic oscillator
potential for the cases $\alpha=\beta=2$.

The Ritz method gives an upper limit $E$ for the ground state energy $E_0$ for the  the fractional stationary Schr\"odinger equation and is given by the expectation values:
\begin{equation}
  \label{ch245Ritz}
  E_0 \leq E = {\braket{\Psi|H^{\alpha,\beta}|\Psi} \over\braket{\Psi | \Psi}}=
   {\braket{\Psi|T^{\alpha/2}+V^{\beta/2}|\Psi} \over\braket{\Psi | \Psi}}
\end{equation}
with a trial function $\Psi$ of the general form
\begin{equation}
  \Psi = \sum_{i=0}^n c_i \psi_i
\end{equation}
We will consider two coordinate systems, namely Cartesian and polar and the two possible
operator sequences according to Riemann and Caputo. 

First we define the covariant non-localization operators with proper normalization $N$ and then calculate the expectation values for norm $\braket{\Psi | \Psi}$, potential energy  $\braket{\Psi|V^{\beta/2}|\Psi}$ and 
kinetic energy $\braket{\Psi|T^{\alpha/2}|\Psi}$ for a trial function $\ket{\Psi}$:

Using Cartesian coordinates on $R^2$ we obtain for the non-localization operator $\tilde{\mathcal{G}}$:
\begin{equation}
\tilde{\mathcal{G}}_c \ast = \frac{1}{N}  \int_{-\infty}^{\infty} dh_x  \int_{-\infty}^{\infty}  dh_y  \, w(h_x,h_y) \cdot  \qquad \qquad   0 \leq \alpha \leq 2 
\end{equation}
Equivalently, introducing polar coordinates on $R^2$ and $dV=\sqrt{g}= h_r$:
\begin{equation}
\tilde{\mathcal{G}}_p \ast = \frac{1}{N}  \int_{0}^{\infty} dh_r \,h_r \int _0^{2 \pi} dh_{\phi}  \, w(h_r,h_{\phi}) \cdot  \qquad \qquad  0 \leq \alpha \leq 2 
\end{equation}
In both cases, we use  a covariant weight $w$.

The norm $N$ is determined by the requirement that the eigenvalue spectrum or Fourier transform  of $\tilde{\mathcal{G}}$ with the covariant Riesz weight (\ref{ch24501}) should yield:
\begin{equation}
  (\mathcal{F} \ast \tilde{\mathcal{G}}) (k) =  (\mathcal{F} \ast \tilde{\mathcal{G}}_{c}) (k) = (\mathcal{F} \ast \tilde{\mathcal{G}}_{p}) (k) = |k|^{\alpha-2} 
\end{equation}
which results in: 
\begin{equation}
\label{ch27norm2}
  N = 2^{2-\alpha} \pi {\Gamma(1-\alpha/2) \over \Gamma(\alpha/2)}
\end{equation}

We interpret the operator $\Delta^{\alpha/2}$ in (\ref{ch245H}) as a fractional extension of the classic Laplace operator (\ref{ch260LL}) and will apply our 
proposed formalism to the kinetic part $T^{\alpha/2}$ of the Hamiltonian in the following: 

With the covariant Riesz weight (\ref{ch24501}), the fractional extension of the local Laplace operator is given using the Caputo-like sequence (\ref{ch245C}):
\begin{equation}
  \label{ch245rieS}
  \Delta^{\alpha/2}_C  =   { \Gamma(\alpha/2) \over 2^{2-\alpha}  \Gamma(1-\alpha/2)} 
   \int_{0}^{\infty} dh_r \,h_r^{1-\alpha} \int _0^{2 \pi} dh_{\phi} \,  \Delta     
\end{equation}
or, using the Riemann-type sequence (\ref{ch245R}):
\begin{equation}
  \label{ch245rieS}
  \Delta^{\alpha/2}_R  =    { \Gamma(\alpha/2) \over 2^{2-\alpha}  \Gamma(1-\alpha/2)} \,
  \Delta  \int_{0}^{\infty} dh_r \,h_r^{1-\alpha} \int _0^{2 \pi} dh_{\phi}    
\end{equation}
We will evaluate (\ref{ch245Ritz}) with a  rotationally invariant Gaussian trial function:
\begin{equation}
  \label{ch245gauss}
  \ket{\Psi} =  e^{-{1 \over 2} \kappa^2 (x^2+  y^2)} = e^{-{1 \over 2} \kappa^2 r^2}
\end{equation}
where  $1/\kappa$ is a measure for the width of the Gaussian. 

\noindent
For this trial function, we obtain for the norm
\begin{eqnarray}
  \label{ch245norm}
  \braket{\Psi |\Psi} &=&  \pi \kappa^{-2} 
\end{eqnarray}
and for the potential energy expectation value 
\begin{eqnarray}
  \label{ch245pot}
  \braket{\Psi | V^{\beta/2} |\Psi} &=&  \braket{\Psi | {1 \over 2} (x^2+y^2)^{\beta/2} |\Psi}  = 
  \braket{\Psi | {1 \over 2} r^{\,\beta} |\Psi} = 
   {1 \over 2} \pi \kappa^{-\beta-2}\Gamma(1 + \beta/2) \nonumber \\
   & & 
\end{eqnarray}

To evaluate the kinetic energy term in (\ref{ch245Ritz}) we will discuss both, the Riemann-type and the Caputo-type fractional extension of the Laplace operator respectively. 

Working with the Riemann-type sequence (\ref{ch245R}) we first obtain for 
$\tilde{\mathcal{G}} \ket{\Psi}$ in Cartesian and polar coordinates respectively:
\begin{equation}
  \label{ch245rieS}
      \tilde{\mathcal{G}} \ast \ket{\Psi} = 
      \tilde{\mathcal{G}}_c  \ast\ket{\Psi} = 
       \tilde{\mathcal{G}}_p  \ast\ket{\Psi} =
      2^{\alpha/2-1}  \kappa^{-2 + \alpha} \Gamma(\alpha/2)  L_{\alpha/2 -1}({\kappa^2 r^2 \over 2 }) \ket{\Psi} 
\end{equation}
where $L_q(z)$ is the Laguerre polynomial of order $q$. 

In a second step we  apply the local Laplace operator $\Delta$, either in Cartesian  $\Delta_c$ and polar coordinates $\Delta_p$ respectively,  which results in:
\begin{eqnarray}
  \Delta^{\alpha/2}_R \ket{\Psi}  & =& \Delta  \tilde{\mathcal{G}}  \ast \ket{\Psi} \label{ch245rieStep1}
  \\ 
    & =& \Delta_c  \tilde{\mathcal{G}}_p  \ast \ket{\Psi} = \Delta_c  \tilde{\mathcal{G}}_c  \ast \ket{\Psi}  \\
    &=&
  \Delta_p  \tilde{\mathcal{G}}_p  \ast \ket{\Psi} = \Delta_p  \tilde{\mathcal{G}}_c  \ast \ket{\Psi}   \label{ch245rieStep2}
  \\
  &=&     2^{\alpha/2-1}  \kappa^{\alpha} \Gamma(\alpha/2) \times  \label{ch245rieStep3}
  \\
  & &\left(    
        (k^2-2) L_{\alpha/2 - 1}({\kappa^2 r^2 \over 2 }) + 
        (k^2-1) L^1_{\alpha/2 - 2}({\kappa^2 r^2 \over 2 })  
     \right. + \nonumber \\
  \label{ch245rieStep22}
     && \,\,\, \,  \left.    (k^2-1) L^1_{\alpha/2 - 2}({\kappa^2 r^2 \over 2 }) +
       k^2 r^2 L^2_{\alpha/2 - 3}({\kappa^2 r^2 \over 2 }) 
       \right) \ket{\Psi} \nonumber
\end{eqnarray}
where $L_q^p(z)$ gives the generalized Laguerre polynomial order $q$. 

With (\ref{ch245rieStep3}) we have applied our 2-step procedure to obtain the action of the fractional Laplace operator on the Gaussian trial function (\ref{ch245gauss}) according to the Riemann-type sequence (\ref{ch245R}). 

We may now use this result (\ref{ch245rieStep3}) to calculate the expectation value of the kinetic energy term in  (\ref{ch245Ritz}).
We get the remarkably simple final result:
\begin{eqnarray}
  \label{ch245kin}
  \braket{\Psi | T^{\alpha/2} |\Psi} &=&  \braket{\Psi | -{1 \over 2} \Delta^{\alpha/2}_R |\Psi}  
  =  {1 \over 2} \pi \kappa^{\alpha-2}\Gamma(1 + \alpha/2) 
\end{eqnarray}
The ground state energy follows with (\ref{ch245kin}),(\ref{ch245pot}),(\ref{ch245norm}) as:
\begin{equation}
  \label{ch245Ritz2}
  E = {\braket{\Psi|T^{\alpha/2}+V^{\beta/2}|\Psi} \over\braket{\Psi | \Psi}} = {1 \over 2} \kappa^{\alpha}\Gamma(1 + \alpha/2) +
  {1 \over 2} \kappa^{-\beta}\Gamma(1 + \beta/2)
\end{equation}
Minimizing the energy with respect to $\kappa$ by solving the equation 
\begin{equation}
  \label{ch245k1}
 \frac{\partial}{\partial \kappa} E(\kappa) = 0  
\end{equation}
yields the optimum $\kappa_{opt}$:
\begin{equation}
  \label{ch245kopt}
  \kappa_{opt}(\alpha, \beta) =\left( {\beta \over \alpha} {\Gamma(1+\beta/2) \over \Gamma(1+\alpha/2)} 
     \right)^{1 /(\alpha+\beta)} 
\end{equation}
and the optimum energy $E_{opt}(\kappa_{opt})$:
\begin{equation}
  \label{ch245eopt}
  E_{opt}(\alpha, \beta) ={\alpha + \beta \over 2 \alpha \beta} 
  \bigl(\alpha\, \Gamma(1+\alpha/2)\bigr)^{\beta/(\alpha+\beta)}
  \bigl(\beta \,  \Gamma(1+\beta/2 )\bigr)^{\alpha/(\alpha+\beta)}
\end{equation}
with the property $E_{opt}(\alpha, \beta) =E_{opt}(\beta, \alpha)$ as a direct consequence of the particle-wave dualism which also holds for the fractional extension of quantum mechanics.

\begin{figure}[t]
  \begin{center}
  \includegraphics[width=\textwidth]{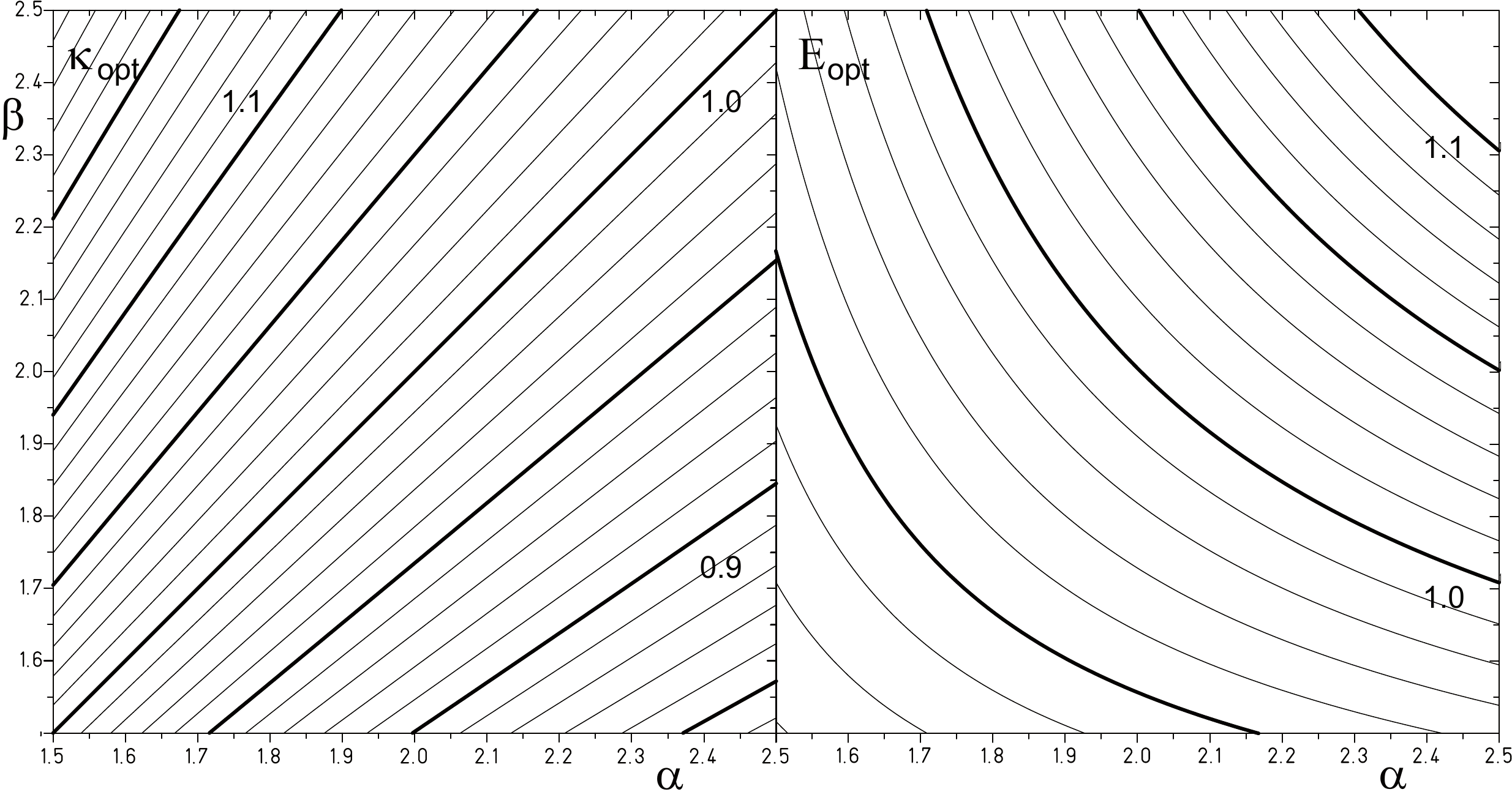}    
  \caption{
    Left: The optimum $\kappa_{opt}$ from (\ref{ch245kopt})
   Right:The optimum $E_{opt}$ from (\ref{ch245eopt})
  }
  \label{fig27001}
  \end{center}
  \end{figure}
  \index{fractional derivative!Riesz}
In Figure \ref{fig27001} we show the corresponding graphs of the optimum $\kappa_{opt}$ and $E_{opt}$.

Let us now compare these results with the Caputo-type sequence (\ref{ch245C}) for a fractional extension of local operators for Cartesian and polar coordinates, respectively for  the local 
$\Delta=\{ \Delta_c, \Delta_p \}$
as well as the non-localization operator $\tilde{\mathcal{G}}=\{ \tilde{\mathcal{G}}_c, \tilde{\mathcal{G}}_p  \}$.

With the same Gaussian trial function $\ket{\Psi}$ we evaluate: 
\begin{eqnarray}
  \label{ch245capStep2}
  \Delta^{\alpha/2}_C \ket{\Psi}  & =& \tilde{\mathcal{G}} \ast \Delta \ket{\Psi} \\
  & =& \tilde{\mathcal{G}}_p \ast \Delta_c\ket{\Psi} =  \tilde{\mathcal{G}}_c \ast\Delta_c \ket{\Psi}  \\
  &=&
  \tilde{\mathcal{G}}_p \ast\Delta_p \ket{\Psi} =   \tilde{\mathcal{G}}_c \ast\Delta_p \ket{\Psi}   \\
    &=& \tilde{\mathcal{G}} \ast \kappa^2(\kappa^2 r^2-2 ) \ket{\Psi} \\
 & =& 
  2^{\alpha/2-1}  \kappa^{\alpha} \Gamma(\alpha/2) \times  \\
& &\left(    
   (\kappa^2-2) L_{\alpha/2 - 1}({\kappa^2 r^2 \over 2 }) + 
   (\kappa^2-1) L^1_{\alpha/2 - 2}({\kappa^2 r^2 \over 2 })  
\right. + \nonumber \\
&& \,\,\, \,  \left.    (\kappa^2-1) L^1_{\alpha/2 - 2}({\kappa^2 r^2 \over 2 }) +
  \kappa^2 r^2 L^2_{\alpha/2 - 3}({\kappa^2 r^2 \over 2 }) 
  \right) \ket{\Psi} \nonumber 
\end{eqnarray}
which is identical with (\ref{ch245rieStep3}) and consequently we obtain the same results
for $\kappa_{opt}$ and $E_{opt}$ as before for the Riemann-type fractional extension sequence for the Gaussian trial function.

We have demonstrated, that either choice of the coordinate system (Cartesian, polar) as well as either choice of the operator sequence (classic operator, non-localization convolution with the Riesz type weight)  calculating the fractional extension of the 2-dimensional Laplace operator leads to identical results.  

This is a strong indication, that the proposed fractional generalization method yields reliable and valid results which are independent of a specifically chosen coordinate set.

In addition, multi-dimensional fractional calculus provides new insights into mathematical and physical phenomena that were not evident in the 1-dimensional case. 

In the next section, we present some examples.

\section{Multi-dimensionality, non-locality and new viewpoints}
We have demonstrated that a covariant fractional extension of a standard local tensor operator may be successfully achieved through a 2-step process.

Additional intriguing aspects arise when extending fractional calculus from one-dimensional to multi-dimensional spaces.

In 1-dimensional fractional calculus, there are only a few distinct definitions of a fractional integral with a given singular weight (see \ref{ch24501}), which differ by setting different integral bounds.

In higher-dimensional spaces, 
the variety of possible regions and topologies increases, as the number of degrees of freedom grows: 

\begin{figure}[t]
  \begin{center}
  \includegraphics[width=\textwidth]{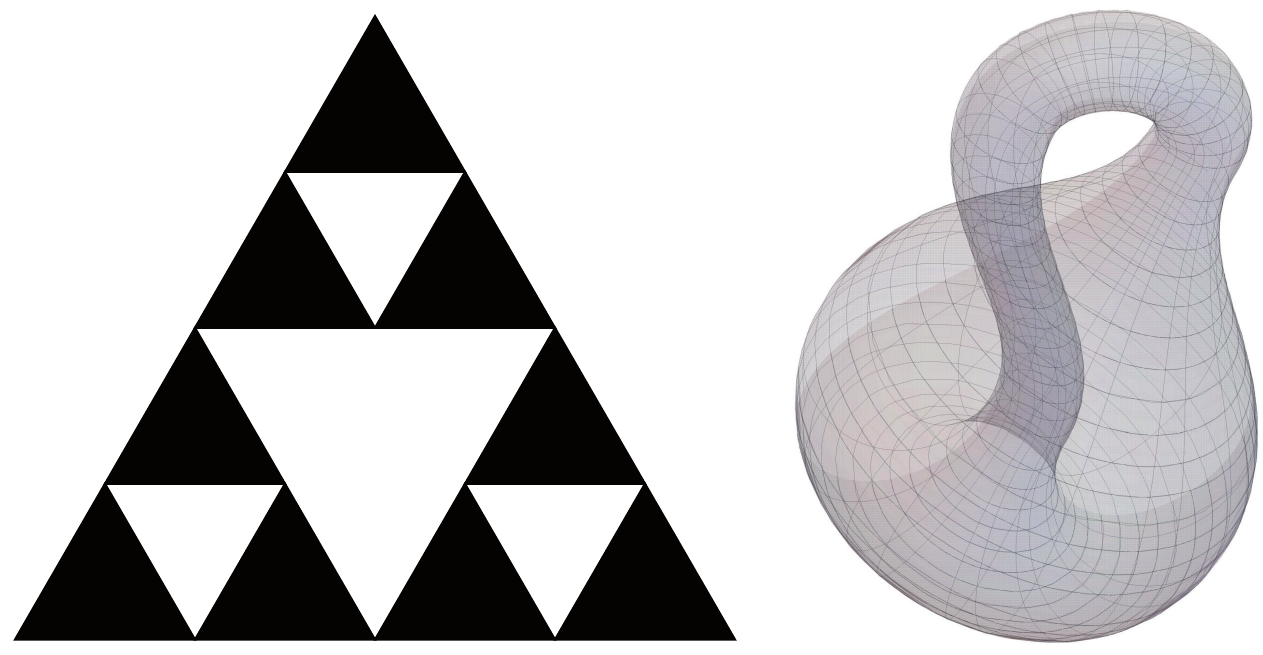}    
  \caption{
  New regions in multi-dimensional fractional calculus include the correct treatment of new possible region types for the non-localization integral  $\tilde{\mathcal{G}}$:   
  On the left: the Sierpi\'nski triangle (with recursion step = 3)  a potential fractal region for the 2-dimensional non-localization operator
  and a typical example for self-similarity. 
  On the right: the Klein bottle, an example of a volume with a non-orientable surface (modified version  of \cite{sen24}).
  }
  \label{fig27002}
  \end{center}
  \end{figure}
  \index{fractional derivative!Riesz}
   Figure \ref{fig27002} illustrates new fascinating regions in multi-dimensional fractional calculus. On the left, we depict a Sierpi\'nski triangle \cite{Sie16} for a specific iteration step, representing one possible realization of a non-connected set of sub-regions for defining a fractional integral, and exhibiting typical fractal properties. \index{Sierpi\'nski triangle}

   \begin{itemize}
    \item
       A notable challenge is integrating over a region like the Klein bottle (shown on the right of Figure \ref{fig27002}), which exemplifies a volume with a non-orientable surface \cite{kle81}.

       \item
       Another aspect of selecting an integration region arises with the covariant extension of the Caputo and Riemann fractional integrals, both of which are typically defined over only a portion of $\mathbb{R}^N$ \cite{pod99}.
   Until now, the coordinate sets for the local operator and the non-local operator could be chosen independently. In the case of the Caputo and Riemann fractional integrals, these coordinate systems are already linked in the 1-dimensional case.

\item
Another important aspect is the use of finite-range weights. 
For example, in three-dimensional spherical coordinates, setting the integral bounds to 
$0\leq h_r\leq r$ instead of $0\leq h_r \leq \infty$ defines a spherical boundary, resulting in a finite-range potential with a sharp cut-off. 
 
 This can be viewed as an extreme case of tempered fractional calculus. In the context of nuclear physics, such a cut-off simulates a finite-range potential, akin to a generalized Yukawa-type potential, which describes the propagation of massive particles.
\end{itemize}

In one dimensional fractional calculus, a classic requirement for valid fractional 
convolution integrals is the restriction to weakly singular weights \cite{die20}.
In higher-dimensional spaces, however, this requirement should be interpreted more broadly.

To clarify this point: 

Consider the 2-dimensional non-localization integral e.g.  in Cartesian coordinates 
$\{p,m\}$ 
\begin{equation}
  \label{ch245two}
  \tilde{\mathcal{G}}_c \ast = 
  \frac{1}{N}  I_2(m) I_1(m,p) \cdot =  
  \frac{1}{N}  \int_{-\infty}^{\infty} dm  \int_{-\infty}^{\infty}  dp  \, w(m,p)   \cdot \qquad \qquad  0 \leq \alpha \leq 2 
\end{equation}
This expression is a sequence of two 1-dimensional integrals, where the inner integral is given by:
\begin{equation}
  \label{ch245twotwo}
  I_1(m,p) \cdot =  \int_{-\infty}^{\infty}  dp  \, w(m,p)  \cdot =  \int_{-\infty}^{\infty}  dp \frac{1}{(p^2+m^2)^{\alpha/2}} \cdot
  \qquad   0 \leq \alpha \leq 2 
\end{equation}
with the covariant Riesz weight  (\ref{ch24501}):   
\begin{equation}
  \label{ch245three}
  w(p,m) = \frac{1}{(p^2+m^2)^{\alpha/2}} \qquad\qquad  \qquad  \qquad  \qquad  \qquad  m^2 \geq 0
\end{equation}
which is singular only for the case $m=0$ and reduces to a non-singular but still 
non-local weight for all $m^2>0$.   

We may conclude that, in higher-dimensional fractional calculus, a non-singular  weight for all 
 $ ds>0 $ is a direct consequence of the covariance requirement.

Extending the coordinate set  to 4-dimensional Minkowski space-time $\{ p_x,p_y,p_z, t, m \}$ 
(\ref{ch245three})  may be interpreted as a key component in the formulation  a generalized Feynman propagator for massive particles. 
\index{Minkowski space-time}
\index{Feynman propagator}
\section{The freedom of choice: The fractional Schr\"odinger operator}
An additional aspect of our proposed approach to fractionalizing a given local operator is the freedom to choose the appropriate non-localization level. To illustrate this, we recall the sequence of operators in the context of the Schrödinger equation. 

Depending on the level of discussion, we may fractionalize only the kinetic energy operator $T$, but also  
the Hamiltonian $H$, or even the Schr\"odinger operator $S$.
\begin{eqnarray}
  T &=& -{\hbar^2 \over 2 m }\Delta \\ 
  \label{Hcalc}
  H &=& T+V = -{\hbar^2 \over 2 m }\Delta + V(x) \\
  \label{ch260S}
  S &=& -{\hbar^2 \over 2 m }\Delta + V(x,t) - i\hbar \partial_t  
\end{eqnarray}
In the previous section, we discussed the non-localization of the kinetic energy operator $T$ only.

What are the consequences of instead applying the non-localization procedure to the Hamiltonian $H$?

From (\ref{Hcalc}) we deduce that the non-localization of $H$ instead of $T$ simply
extends the non-localization procedure now including the potential operator  $V$, 
which results in the fractionalized version of the Hamiltonian $\tilde{\mathcal{H}}_{\textrm{R,C}}(w)$ of either Riemann- 
\begin{eqnarray}
  \label{ch245HR}
  \tilde{\mathcal{H}}_{\textrm{R}}(w)   &=& 
  \tilde{\mathcal{T}}_{\textrm{R}}(w) + \tilde{\mathcal{V}}_{\textrm{R}}(w)  \\
  &=&  H  \cdot \tilde{\mathcal{G}}(w) \ast\\
  &=&   (T + V)  \cdot \tilde{\mathcal{G}}(w) \ast
  \end{eqnarray}
  or Caputo-type 
  \begin{eqnarray}
    \label{ch245HC}
    \tilde{\mathcal{H}}_{\textrm{C}}(w)  
     &=& \tilde{\mathcal{T}}_{\textrm{C}}(w) + \tilde{\mathcal{V}}_{\textrm{C}}(w)  \\
     &=&  \tilde{\mathcal{G}}(w) \ast H  \cdot \\
     &=&  \tilde{\mathcal{G}}(w) \ast (T+V)  \cdot 
    \end{eqnarray}
We are interested in the eigenvalues $E_N$ and the eigenfunctions $\ket{N}$ of the fractionalized
Hamiltonian:
\begin{eqnarray}
  \label{ch245HRC}
  \tilde{\mathcal{H}}_{\textrm{R, C}}(w)\ket{N}   &=& E_N \ket{N}  
  \end{eqnarray}
Interpreting the non-localization procedure as an perturbation of the Hamiltonian $\tilde{\mathcal{H}}_0$ of an unperturbed system, a nice application of perturbation theory follows immediately:

With:
\begin{eqnarray}
\label{ch245HRC}
\tilde{\mathcal{H}}_0 \ket{n}   &=& H \ket{n} =  e_n \ket{n}  
\end{eqnarray}
and expanding eigenvalues $E_N$ and the eigenfunctions $\ket{N}$ in a power series in $\lambda$
\begin{eqnarray}
\label{ch245HRC}
  \ket{N}   &=& \sum_{i=0}^\infty \lambda^i \ket{n^{(i)}} \\
  E_N   &=& \sum_{i=0}^\infty \lambda^i E_N^{(i)}
\end{eqnarray}
with $\ket{n^{(0)}} = \ket{n}$ and
 $E_N^{(0)} = e_n $.
 
 We obtain in lowest order a shifted energy given as the expectation value of the non-localized 
$ \tilde{\mathcal{H}}_{\textrm{R, C}}(w)$ with the corresponding unperturbed eigenstate $\ket{n}$:
\begin{eqnarray}
  \label{ch245HRC}
  E_N^{(1)}   &=& \braket{n| \tilde{\mathcal{H}}_{\textrm{R, C}}(w)|n}
  \end{eqnarray} 

For the Caputo-type $ \tilde{\mathcal{H}}_{\textrm{C}}(w)$ without loss of generality  we may simplify further:
\begin{eqnarray}
  \label{ch245CCC}
  _CE_N^{(1)}  &=& \braket{n| \tilde{\mathcal{H}}_{\textrm{C}}(w)|n} \\
  &=&  \braket{n| \tilde{\mathcal{G}}(w) \ast (T+V)|n} \\
  &=& e_n \braket{n| \tilde{\mathcal{G}}(w) \ast|n} 
\end{eqnarray}  
We obtain the remarkable result, that the energy shift for the Caputo-type fractional Hamiltonian
is given in lowest order perturbation theory 
as the expectation value of the non-localization operator acting on the local eigenstates.

As an example let us calculate the spectrum of the 1-dimensional fractional harmonic oscillator in lowest
order perturbation theory.

The eigenfunctions and the eigenvalue spectrum of the 1-dimensional harmonic oscillator in space representation are given as:
\begin{eqnarray}
  \label{ch245HRC}
  H \ket{n}   &=& (-{1 \over 2} \partial^2_x + {1 \over 2} x^2)\ket{n} = e_n \ket{n}\\
  e_n &=& n+{1 \over 2}\\
  \ket{n} &=& \frac{1}{\sqrt{2^n n! \sqrt{\pi}}} e^{-{1 \over 2}x^2} H_n(x) = \psi_n(x)
\end{eqnarray} 
We evaluate using the Riesz weight and the norm $N=1/(2 \Gamma(1-\alpha))$
\begin{eqnarray}
  \label{ch245CC1}
 \braket{n| \tilde{\mathcal{G}}(w) \ast|n} 
 &=&\frac{1}{N}
 \int_{-\infty}^{+\infty}dx \,  \psi_n(x) \int_{-\infty}^{+\infty}  dh  {1 \over |h|^\alpha}\psi_n(x+h)  \\
 &=&\frac{1}{N}
 \int_{-\infty}^{+\infty}    {dh \over |h|^\alpha} \int_{-\infty}^{+\infty}dx \,  \psi_n(x) \psi_n(x+h)  \\
 \label{ch245CCG1}
 &=&\frac{2}{N}
 \int_{0}^{+\infty}  {dh \over h^\alpha} \int_{-\infty}^{+\infty}dx \,  \psi_n(x) \psi_n(x+h)  \\
 \label{ch245CCG2}
 &=&\frac{2}{N}
 \int_{0}^{+\infty}  {dh \over h^\alpha} e^{-h^2/4}L_{n}(h^2/2)  
\end{eqnarray}  
(\ref{ch245CCG2}) results from (\ref{ch245CCG1}) since the inner integral in (\ref{ch245CCG1}) is nothing else but a shifted harmonic oscillator or Glauber state
 \cite{cah69}
and $L_{n}(\xi)$ is a Laguerre polynomial of order $n$.
Since the Laguerre polynomials are given explicitly as
\begin{eqnarray}
  \label{ch245CC1}
 L_n(\xi) &=&
 \sum_{j=0}^n 
 \left(\begin{array}{c}
  n \\ j \\
  \end{array} \right)
  {(-1)^j \over j!}\xi^j 
\end{eqnarray} 
we may write explicitly for (\ref{ch245CCG2}) :
\begin{eqnarray}
  \braket{n| \tilde{\mathcal{G}}(w) \ast|n} 
 &=&
 \frac{2}{N} \int_{0}^{+\infty}  {dh \over h^\alpha} e^{-h^2/4}L_{n}(h^2/2)  \\
&=& \frac{2}{N}
\sum_{j=0}^n 
\left(\begin{array}{c}
 n \\ j \\
 \end{array} \right)
  {(-1)^j \over 2^j j!}  \int_{0}^{+\infty}  dh e^{-h^2/4} h^{2 j - \alpha}
\end{eqnarray}  
The integral is a scaled gamma function
\begin{eqnarray}
  \label{ch245CC1}
  \int_{0}^{+\infty}  dh e^{-h^2/4} h^{2 j - \alpha} 
  &=&
  2^{2 j - \alpha} \Gamma(j + \frac{1-\alpha}{2})
\end{eqnarray} 
which leads to  
\begin{eqnarray}
  \braket{n| \tilde{\mathcal{G}}(w) \ast|n} 
 &=&
 \frac{\sqrt{\pi}}{\Gamma(1-\alpha/2)} {_2F_1}(\frac{1-\alpha}{2},-n;1;2)
\end{eqnarray}  
We obtain in first order perturbation expansion  for the energy of the fractional harmonic oscillator
with  Hamilton operator non-localized according to Caputo-type:
\begin{eqnarray}
  \label{ch245HRC}
    _CE_N^{(1)}(\alpha, n) &=& (n+\frac{1}{2})\frac{\sqrt{\pi}}{\Gamma(1-\alpha/2)} {_2F_1}(\frac{1-\alpha}{2},-n;1;2)
  \end{eqnarray} 
Let us recall the WKB approximation of the energy levels of the harmonic oscillator\cite{las02}:  
\begin{eqnarray}
  \label{ch245WKB}
    E_{WKB}(\alpha, \beta, n) &=& \frac{1}{2} 
    \left(
    (n+\frac{1}{2}) 
    \frac{ \pi \beta }{2 B(1/\alpha, 1/\beta+1)}
    \right)^{\frac{\alpha \beta}{\alpha + \beta}}
  \end{eqnarray} 
with the beta function $B(x,y)$.  

\begin{figure}[t]
  \begin{center}
  \includegraphics[width=\textwidth]{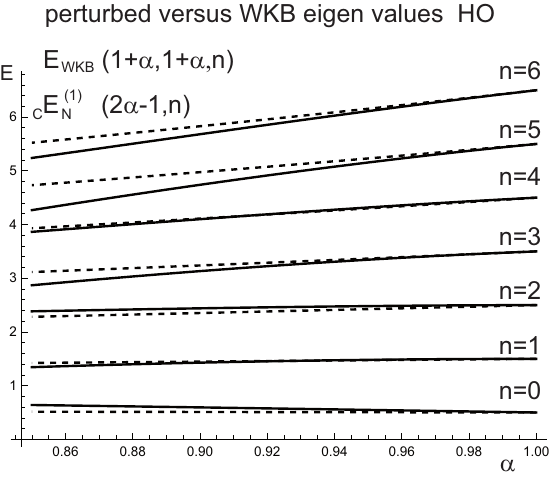}    
  \caption{
    Comparison for the lowest $n=0,...,6$ energy levels of the fractional harmonic oscillator in lowest order perturbation correction  $_CE_N^{(1)}(2 \alpha-1, n)$ (thick lines ) according to (\ref{ch245HRC}) with 
    the WKB approximation $ E_{WKB}(1+\alpha, 1+\alpha, n) $ (dashed lines) according to  (\ref{ch245WKB}). For $\alpha=1$ both approximations reproduce the equidistant level spectrum of the standard harmonic oscillator exacltly.
  }
  \label{fig27003}
  \end{center}
  \end{figure}

We obtain the remarkable relation valid near $\alpha=1$:
\begin{eqnarray}
  \label{ch245W3}
    _CE_N^{(1)}(2 \alpha-1, n) &\approx& E_{WKB}(1+\alpha, 1+\alpha, n) \qquad \alpha = 1 -\epsilon
  \end{eqnarray}   
In Figure \ref{fig27003} we compare both approximations for the lowest eigenvalues of the fractional
harmonic oscillator in the range $0.85 \leq \alpha \leq 1$.    

  \index{fractional derivative!Riesz}

In the lowest-order perturbation expansion, we may interpret the non-localization of the potential operator as a fractional integral, which transforms the harmonic oscillator potential into a more general power-law potential.

To ensure that a fractional extension of quantum mechanics treats coordinates and conjugate momenta symmetrically—preserving particle-wave duality and providing a description independent of whether we work in coordinate or momentum space—we apply the following transformation:
\begin{equation}
\label{c8q}
\{x^n, {d^n \over dx^n }\} \rightarrow \{x^\alpha, {d^\alpha \over dx^\alpha }\}
\end{equation}
Based on this, we conclude that a proper fractional analogue of classic quantum mechanics should fractionalize not only the kinetic energy operator  $T$ but the entire Hamiltonian $H = T+V$ including
the coordinate-dependent potential energy term. 
\section{Minimal coupling of an electromagnetic field}
So far, we have discussed the non-localization of the kinetic energy operator
 $\hat{T}$ and the Hamiltonian
$\hat{H}=\hat{T}+\hat{V}$, which represents one way of realizing the step from the free Schrödinger equation to a wave equation that includes a potential.

However, there is another significant approach worth considering in our method of fractionalizing local operators, when the Schrödinger operator $\hat{S}=\hat{H}-\hat{E}$ (\ref{ch260S}) is fractionalized. This involves the covariant coupling of external charges \cite{sak67, mes68, gre01b, sre07, kha22}. 

The requirement of local gauge invariance and the resulting minimal coupling scheme - such as for the electromagnetic field in the Schrödinger equation - may also provide a useful mechanism in fractional quantum field theories.

In essence, to preserve local gauge invariance, the momentum and energy operators are modified as follows:
\begin{eqnarray}
  \label{ch245ppp}
  \hat{p} &=& -i \hbar \nabla \rightarrow -i \hbar \nabla - q \vec{A} \\
  \hat{E} &=& i \hbar \partial_t \rightarrow i \hbar \partial_t - q V 
\end{eqnarray}   
where $\vec{A}$ is the electromagnetic vector potential, and $V$ the scalar potential. 

This leads to a 
modified Schr\"odinger operator $\hat{S}^\ast$ in the presence of an electromagnetic potential:  
\begin{eqnarray}
  \label{ch245ppp}
  \hat{S}^\ast = \frac{1}{2 m} (-i \hbar \nabla - q \vec{A})^2 - (i \hbar \partial_t - q V ) 
\end{eqnarray}   
which now may be fractionalized in the next step to obtain the fractionalized version of modified Schr\"odinger operator 
$\tilde{\mathcal{S}}_{\textrm{R,C}}(w)$ of either Riemann- 
\begin{eqnarray}
  \label{ch245SR}
  \tilde{\mathcal{S}}^\ast_{\textrm{R}}(w)   &=& 
  \tilde{\mathcal{H}}_{\textrm{R}}(w) - \tilde{\mathcal{E}}_{\textrm{R}}(w)  \\
  &=&  (H-E)  \cdot \tilde{\mathcal{G}}(w) \ast
  \end{eqnarray}
  or Caputo-type 
  \begin{eqnarray}
    \label{ch245SC}
    \tilde{\mathcal{S}}^\ast_{\textrm{C}}(w)  
     &=& \tilde{\mathcal{H}}_{\textrm{C}}(w) - \tilde{\mathcal{E}}_{\textrm{C}}(w)  \\
     &=&  \tilde{\mathcal{G}}(w) \ast (H-E)  \cdot  
\end{eqnarray}
\section{Conclusions}
We have proposed a covariant transition from local to fractional operators, which is based  
on the requirement of covariance for all constituents involved: the local operator, the weight and the non-localization (convolution) 
integral on $\mathbb{R}^N$ leading to the fractional extended operator.
It is a two-step procedure, which may be directly applied for a wide range of weights, weakly singular and non-singular as well.  

We have applied the proposed procedure to calculate an approximation of the ground state energy of the fractional 2-dimensional harmonic oscillator and obtained an analytic formula using the Ritz variation principle.

In addition, we gave a new analytic result for the energy level spectrum of the one-dimensional fractional harmonic oscillator in lowest order perturbation theory.
       
The proposed non-localization procedure on an appropriately chosen metric 
establishes a connection between fractional calculus and  generalized field theories.
This opens new and promising insights and applications in e.g. 
cosmology or particle physics.

\section{Acknowledgement}
We thank A. Friedrich for useful discussions.



\end{document}